# Somatosensory and motor contributions to emotion representation

Marianne Reddan,[1] Luke Chang,[2] Phil Kragel,[3] Tor Wager[2]

[1]Department of Psychiatry and Behavioral Sciences, Albert Einstein College of Medicine, Bronx, NY; [2]Department of Psychological and Brain Sciences, Dartmouth College, Hanover, NH, [3]Department of Psychology, Emory University, Atlanta, GA

## Abstract

Emotion is often described as something people 'feel' in their bodies. Embodied emotion theorists propose that this connection is not purely linguistic; perceiving an emotion may require somatosensory and motor reexperiencing. However, it remains unclear whether self-reports of emotion-related bodily sensations (i.e., 'lump in my throat') are related to neural simulations of bodily action and sensation or whether they can be explained by cognitive appraisals or the visual features of socioemotional signals. To investigate this, participants (N = 21) were shown arousing emotional images that varied in valence, complexity, and content while undergoing fMRI scans. Participants then rated the images on a set of emotion appraisal scales and indicated where, on a body map, they experienced sensation in response to the image. To derive normative models of responses on these scales, a separate larger online sample online (N = 56 - 128) also rated these images. Representational similarity analysis (RSA) was used to compare the emotional content in the body maps with appraisals and visual features. A pairwise distance matrix between the body maps generated for each stimulus was then used in a whole brain voxel-wise searchlight analysis to identify brain regions which reflect the representational geometry of embodied emotion. This analysis revealed a network including bilateral primary somatosensory and motor cortices, precuneus, insula, and medial prefrontal cortex. The results of this study suggest that the relationship between emotion and the body is not purely conceptual: It is supported by sensorimotor cortical activations.

## Introduction

Emotion theorists, dating back to Darwin have argued that our bodies are not just tools for expression and action, but instead, active participants in an emotional experience (Darwin, 1872). For example, William James proposed that the emotion itself arises only after the brain recognizes some change in the body's physiology (James, 1884). Indeed, emotions incite profuse bodily changes, including, but not limited to physiological arousal, visceral sensations, facial contortions, gestures, and action readiness (de Gelder, 2006; Ekman, 1993; Levenson, 2014). When people describe their emotions they often refer to feelings in their bodies (Damasio et al., 2000). For example, people may recount bodily sensations such as rising temperature, furrowed brows, and even a balled-up fist when recalling feelings of rage (Niedenthal, 2007). Our language echoes this connection (Barrett et al., 2007). Common phrases used to express emotion have bodily referents: We experience gut feelings, a lump in the throat, hot headedness, shivers down the spine, and a sunken heart. Indeed, when asked to indicate on a silhouette of the human body where one "feels" an emotion, people produce statistically separable topographic bodily sensation maps for discrete complex and basic emotional states, and these maps are consistent across Eastern and Western cultures (L. Nummenmaa, Glerean, Hari, & Hietanen, 2013; Lauri Nummenmaa, Hari, Hietanen, & Glerean, 2018). Despite this profound collection of behavioral evidence for the embodiment of emotion, it remains unknown whether this relationship between brain and body is purely conceptual, or if it has a neural basis. Does the neural construction of emotion include activation in cortical regions specialized for bodily sensation and action? This investigation aims to provide a neural link for the embodiment of emotion by analyzing the representational similarity between self-reported topographic body maps of emotion and neural activation in sensorimotor regions of interest.

Embodied emotion is a subcomponent of grounded cognition theory, which argues that all cognition is supported by simulation, and that both concrete and abstract knowledge representations retain some modal component (i.e., of perception (the senses), action (movement and proprioception), or introspection (mental states and affect); Barsalou, 2007). In this framework, emotions are grounded in bodily sensations, that is, an emotional experience is dependent upon neural simulations within sensorimotor circuitry (Wilson-Mendenhall et al., 2012). Neuroimaging and lesion studies support this proposal. For example, Kragel and Labar (2016) showed that the perception of happy faces, which have more mouth-based features, activate areas of the primary somatosensory cortex representing the lips, tongue, jaw and mouth, while the perception of fearful faces, which have more eye-based features, activate areas that represent the brow. Furthermore, patients with right ventral primary and secondary somatosensory cortex lesions demonstrate impaired recognition of emotional facial expressions, despite having intact visual processing streams (Adolphs et al., 2000). Drawing across this evidence, embodied emotion theory purports that somatosensory and motor simulations are critical to the expression and recognition of emotional states.

To test whether somatosensory and motor simulations are related to self-reported emotion-related sensation in one's body, this experiment interrogates the similarity between self-reported topographic maps of bodily emotion and neural representations within sensorimotor and perceptual cortices using neuroimaging and multivariate analyses. In this experiment, in-lab participants viewed emotional images that varied in valence, complexity, and content, while high resolution images of their brains were collected in an fMRI scanner. Appraisals of these images were collected in separate larger online sample (N = 56 - 128), so that neural activation was not compromised by experimental demands related to cognitive appraisals or button responses. We also did this so that we could construct normative models of emotion based on the stimuli. Representational similarity analysis (RSA; Nili et al., 2014) was used to compare the representation of emotional content at three levels of analysis: embodiment self-reports, cognitive appraisals, and neural activity. In addition, high-level visual features of the stimuli were extracted from the last layer of AlexNet (Krizhevsky, Sutskever, & Hinton, 2012). Through this analysis we were able to test whether embodiment self-reports were unique from or equally explainable by appraisal patterns or high-level visual features of the stimuli (i.e., scenes or objects).

## Materials and Method

*Participants*. Participants (N=21) were recruited from the University of Colorado-Boulder and the surrounding community (10 Female, Average Age = 24.5, right-handed). Participants did not meet DSM V criteria for psychological disorders, had no life history of head trauma, nor any contraindications for the MR environment, meaning their bodies were free of ferromagnetic substances.

### Study Design

*In Lab*. Picture stimuli were selected from the International Affective Picture System System (IAPS) and the Geneva Affective Picture Database (GAPED) using published normative arousal ratings, so that each image was selected to be highly arousing and so that there were an equal number of positive and negative valenced images (Bertron et al., 1997; Dan-Glauser & Scherer, 2011; Libkuman, Otani, Kern, Viger, & Novak, 2007; Mikels et al., 2005). 112 images were used in this experiment. Images were presented while the subject was lying in the scanner. Image presentation lasted 4-seconds, with a randomized jittered inter-trial-interval of 2 to 12-seconds (average ISI time per run = 4-sec). The image presentations were divided into two 7.5 minute runs, where the first 56 images were presented in a randomized order, followed by a random presentation of the remaining 56 images. The MATLAB Psychtoolbox package was used for stimulus presentation. Scans were acquired using a multiband imaging protocol with a 460 ms TR on a Siemens Allegra (3 Tesla). Distortion correction scans were collected for multiband image preprocessing.

*Online.* In order to control for motor-response tendencies and meta-awareness of one's appraisals, appraisal ratings were collected online with a new cohort (N= 56 - 128) on mechanical Turk. Participants were given a random set of images, the number of images dependent upon how much time they were willing to spend on the experiment. 53 participants completed all ratings of all images and 75 participants partially completed the ratings. The 112 images were rated on 19 predetermined appraisal dimensions (see Table 1). Ratings were given on a visual analogue scale with the anchors (NOT AT ALL, NEUTRAL, and EXTREMELY SO). Participants could mark anywhere on the scale and submit their rating with a click. Ratings were encoded using a 1 - 100 scale.

**Table 1. Image Appraisal Dimensions**

| *Appraisal Dimensions* |
|---|
| Rate the degree to which someone or something is SUFFERING in this image. |
| How strongly do you want to AVOID seeing this image? |
| How ATTENTION-GRABBING is this image? |
| Rate how much this image makes you feel DISGUST. |
| Rate how much this image makes you feel ANGER. |
| Rate how much this image makes you feel SADNESS. |
| Rate how much this image makes you feel FEAR. |
| Rate how much this image makes you feel JOY. |
| Rate how much this image makes you feel SURPRISE. |
| Much does this image make you think of CONTAMINATION or DISEASE? |
| How physically THREATENED do you feel by this image? |
| How PLEASANT is this image? |
| How IMMORAL are the events depicted in this image? |
| Rate how much EMPATHY you feel for the people or animals in this image? |
| How much do you sense that a STORY is beginning to unfold in this image? |
| How RELEVANT is the scene of this image to your life? |
| How SOCIALLY ACCEPTABLE are the events in this image? |
| Rate on this scale of 0 to 60 seconds, when you think you felt emotion in response to seeing the image at your left. |

Participants were also asked to "indicate on the body map below (Figure 1), where on your body, if at all, you feel the emotions elicited by this image." Participants were given the opportunity to submit blank body maps if they did not experience any sensation in their body in response to the image. Body map self-reports were encoded in a binary fashion (on or off, by pixel). Similar ratings were collected in the in-lab sample after the scan, however, due to technical errors, data were lost (N = 7). Therefore, we used average ratings from the large online population to construct models of the brain data. To justify this decision we correlated the responses of the surviving in-lab survey and the online survey. There was high fidelity, therefore this approach is justified.

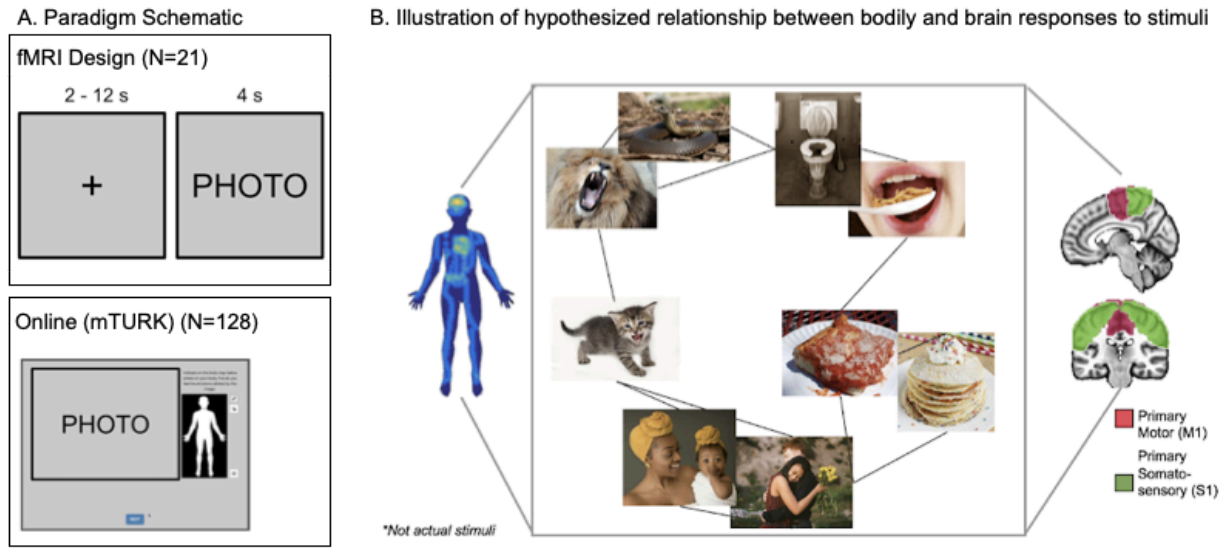

**Figure 1. Experiment Design.** *A. Paradigm Schematic.* Left. Schematic of the paradigm participants (N = 21) experienced in the scanner. 112 highly arousing emotional images from IAPS and GAPED databases were displayed. The same set of images were shown to a new sample (N = 128) online on Amazon's Mechanical Turk, and participants rated these images on 18 different appraisal dimensions. Participants were also asked to color on a body map, where, if at all, they experience a response to the image in their body. Body map reports were encoded in a binary fashion (on or off, by pixel). *B. Hypothesis.* We hypothesized that if emotions are embodied, the representational geometry of emotion in the self-reported body maps would be conserved in the primary somatosensory and motor cortices of the brain.

## Analysis

*Body Map Preprocessing.* Single trial binary self-reported bodily activation maps were reduced in dimensionality to 200 x 100 pixels and then smoothed using a radial dilation (5 pixel diameter). Radial dilation allowed for graded activation in pixels, in a range of 0 to 1. Body maps were vectorized, meaning each pixel activation value was stacked into a single vector in a meaningful order, for further analysis.

*Neuroimaging Acquisition.* Gradient-echo echo-planar imaging BOLD-fMRI was performed on a 3 Tesla Siemens MRI scanner (Siemens Healthcare). Functional images were acquired using Multiband EPI sequence: echo time = 30 ms, repetition time = 460 ms, flip angle = 44°, number of slices = 80, slice orientation = coronal, phase encoding = h > f, voxel size = 1.6 × 1.6 × 2.0 mm, gap between slices = 0 mm, field of view = 191 × 191 mm$^2$, Multi-band acceleration factor = 8; echo spacing = 0.72 ms, bandwidth = 1,724 Hz per pixel, partial Fourier in the phase encode direction: 7/8.

Structural images were acquired using a single shot T1 MPRAGE sequence: echo time = 2.01 ms, repetition time = 2.4 s, flip angle = 8°, number of slices = 224, slice orientation = sagittal, voxel size = 0.8 mm isotropic, gap between slices = 0 mm, field of view = 256 × 256 mm$^2$, GRAPPA acceleration factor = 2; echo spacing = 7.4 ms, bandwidth = 240 Hz per pixel.

*MRI preprocessing.* Multiband brain imaging data were preprocessed following procedures used in the Human Connectome Project (Glasser et al., 2013). This approach includes distortion correction, spatial realignment based on translation (in the transverse, sagittal, and coronal planes) and rotation (roll, pitch, and yaw), spatial normalization to MNI152 space using T1 data, and smoothing using a 6mm FWHM Gaussian kernel. Preprocessing was completed using the Mind Research Network's Auto-Analysis software (Bockholt et al., 2010).

*MRI analysis.* Preprocessed fMRI data were analyzed using general linear models with SPM 8 software (Wellcome Trust Centre for Neuroimaging, UK). Separate models were estimated for each participant that included: (1) a regressor for every image presented to subjects, modeled as a 4s boxcar convolved with the canonical hemodynamic response function of SPM, (2) 24 motion covariates from spatial realignment (i.e., translation in x, y, and z dimensions; roll, pitch, and yaw; and their first and second order temporal derivatives), (3) nuisance regressors specifying outlier timepoints, or 'spikes', that had large deviations in whole-brain BOLD signal, and (4) constant terms to model the mean of each imaging session.

*Model Construction.* The online ratings were used to construct three models of representational similarity across items (images) based on: (1) experienced body locations, (2) appraisal patterns, and (3) image features. The embodiment model consisted of a pairwise distance matrix for every pair of images (Figure 1A). Similarly, the appraisal model was a pairwise distance matrix based on the average 18 appraisal ratings for each image. The visual image feature model was created by applying the object-recognition neural network, AlexNet (Krizhevsky, Sutskever, & Hinton, 2012), to our stimuli and extracting activation in the last fully connected layer, which contains a single unit for each of 1,000 object categories.

*Embodiment Model.* The body maps were vectorized and averaged for each stimulus image. The number of body maps varies from 53 to 128 online user responses due to partial completion of the task. A representational dissimilarity matrix (RDM) was constructed by taking the one minus the Pearson's correlation of average body map vectors for every pair of stimulus images, so that 112 x 112 confusion matrix resulted. The RDM was normalized using a Fisher R to Z transformation. This RDM is a model for embodied emotion because it reflects how sensations in the body are represented for each image relative to one another and it can be used to compare how information is organized in the brain in response to the images. To better understand this model of embodied emotion, we created a descriptive dendrogram of the distance matrix using ward linkage. The dendrogram revealed that reports of embodiment are organized according to the content of the stimulus images. For example, appetizing food items were more similar to one another and clustered together, as did immoral and threatening scenes, and arousing scenes like romantic couples and erotica (Figure 2A). Next this model was applied in a searchlight analysis (radius 4 voxels, total 100) of the whole brain (Su, Fonteneau, Marslen-Wilson, & Kriegeskorte, 2012).

### a. Schematic of the Embodiment Model Construction

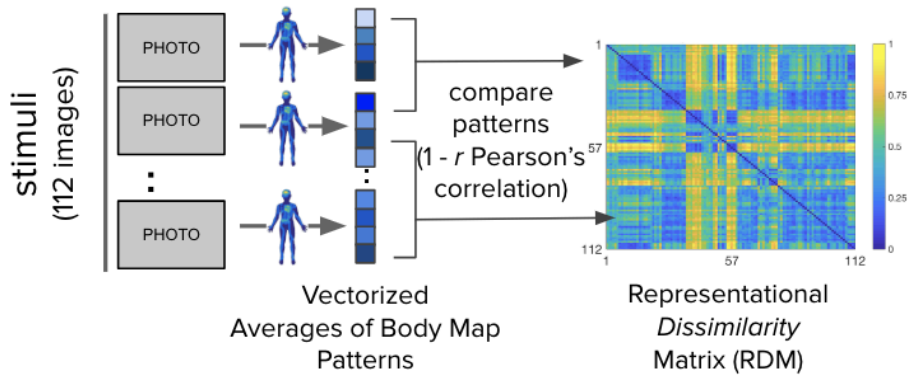

### b. Embodiment Model Organization

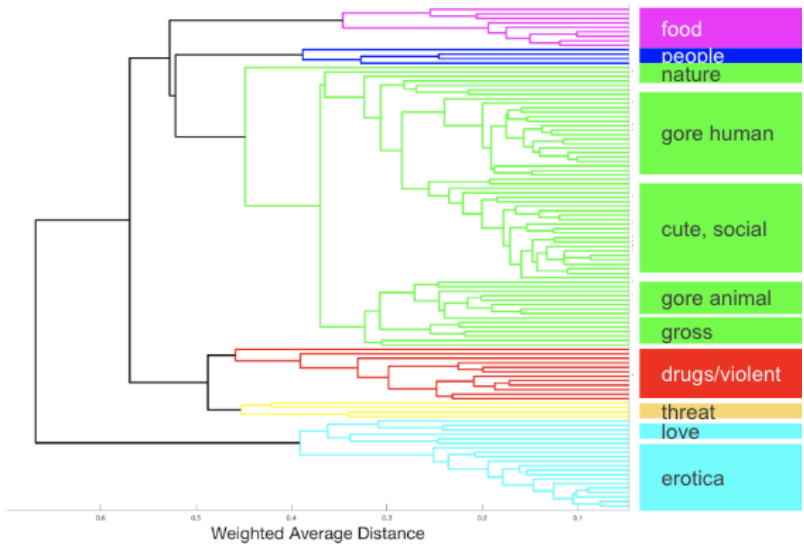

**Figure 2. Embodiment Model of Emotion.** *A. Schematic of the Embodiment Model Construction.* The body maps were vectorized and averaged for each stimulus image. A representational dissimilarity matrix (RDM) was constructed and normalized using a Fisher R to Z transformation. *B. Embodiment Model Organization.* This RDM is a model for embodied emotion because it reflects how sensations in the body are represented for each image relative to one another and it can be used to compare how information is organized in the brain in response to the images. To better understand this model of embodied emotion, we created a descriptive dendrogram of the distance matrix. The dendrogram revealed that reports of embodiment are organized according to the content of the stimulus images. For example, appetizing food items were more similar to one another and clustered together, as did immoral and threatening scenes, and arousing scenes like romantic couples and erotica.

*Appraisal Model.* The values of the appraisal ratings (Table 1) averaged for each stimulus image and then a vector of appraisals was constructed for each image (18 x1). An appraisal model RDM was constructed by taking the one minus the Pearson's correlation of average body map vectors for every pair of stimulus images, so that 112 x 112 confusion matrix resulted. The RDM was normalized using a Fisher R to Z transformation.

*Visual Features Model.* To construct the model of image features, that is the objects and scenes depicted, we used the convolutional neural network for object recognition known as AlexNet. We retrained AlexNet on our stimulus images, and then extracted the weights from the 3$^{rd}$ (last) fully connected layer (the object recognition layer) and constructed the 112 x 112 RDM. The RDM was normalized using a Fisher R to Z transformation.

*Searchlight Analysis.* Local fMRI representational similarity matrices were then regressed on each of these models in whole brain searchlight analyses (radius 4 voxels, total 100) to identify brain regions whose activity reflects the representational geometry of (1) embodied emotion, (2) cognitive appraisals, and (3) stimulus-specific image features. This analysis is similar to that in Su, Fonteneau, Marslen-Wilson, & Kriegeskorte (2012). These brain maps were then compared using support vector machines.

## Results

*Searchlight.* The embodiment searchlight analysis results were thresholded at $p < 0.01$ (uncorrected; results did not survive FDR correction) and revealed a neural network including bilateral primary somatosensory and motor cortices, precuneus, insula, and medial prefrontal cortex (Figure 3). A priori anatomical masks of the primary somatosensory and motor cortices from the Harvard-Oxford Atlas were applied to the thresholded results map in order to confirm that activity was found in our hypothesized regions. The appraisal searchlight results were thresholded (FDR *q* < 0.05) and revealed that activity in the prefrontal cortex were primarily related to the appraisal model. The visual features searchlight results were also thresholded (FDR *q* < 0.05) and revealed that activity in the lateral occipital cortex, a region important for object recognition (Grill-Spector, Kourtzi, & Kanwisher, 2001), was primarily related to the visual features model.

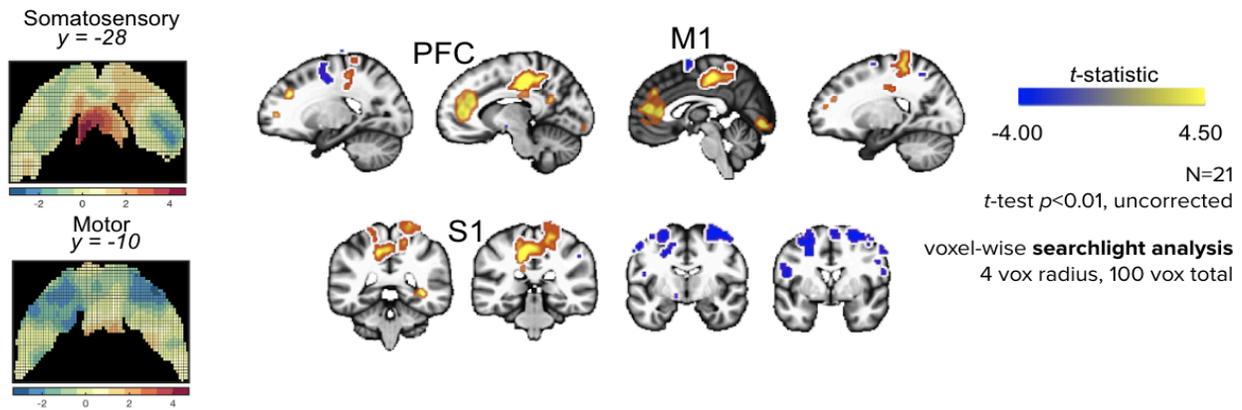

**Figure 3. Embodiment Model Searchlight Results.** The embodiment searchlight analysis results were thresholded at p < 0.01 (uncorrected) and revealed a neural network including bilateral primary somatosensory and motor cortices, precuneus, insula, and medial prefrontal cortex. A priori regions of interest include S1 and M1 which are outlined in white in the thresholded plot and plotted again unthresholded to the left.

*Model Comparison*. A linear binary support vector machine (SVM, C = 1) trained on individual subjects brain maps from the searchlight analysis showed that the representational basis for embodiment and appraisals were distinct from one another (leave-one-out cross-validated accuracy = 85.7%, p <0.0001; Figure 4A), and that the representational basis for embodiment and image features were also distinct from one another (leave-one-out cross-validated accuracy of 97.6%, p <0.001; Figure 4B).

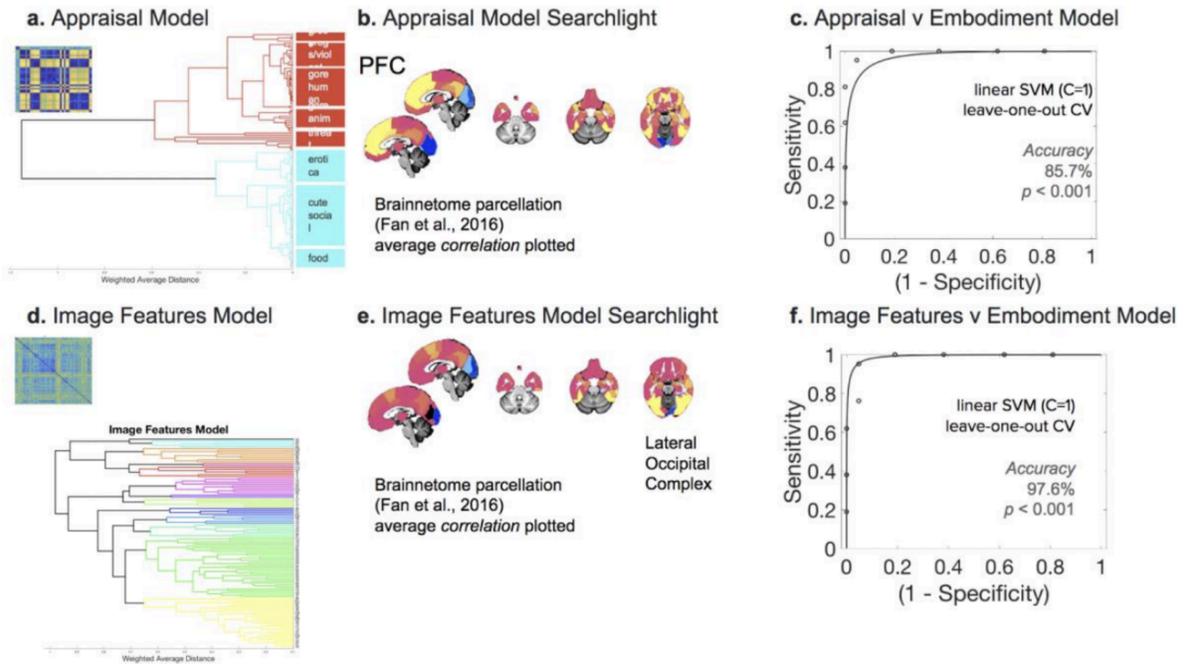

**Figure 4. Model Comparisons.** *A. Appraisal Model.* The appraisal model exhibited a structure consistent with representations in PFC, and a linear SVM (C = 1) on the searchlight results showed that the representational basis for embodiment and appraisals were distinct from one another (leave-one-out cross-validated accuracy = 85.7%, $p < 0.0001$). *B. Image Features Model.* The image features model was structured most consistently with representations in the lateral occipital cortex, a linear SVM (C = 1) showed that the representational basis for embodiment and image features were also distinct from one another (leave-one-out cross-validated accuracy of 97.6%, $p < 0.001$).

## Discussion

The results of this study suggest that the relationship between emotion and the body is grounded in sensorimotor representations: The neural activity related to embodied emotion is different from that related to cognitive appraisals and the visual features of the stimuli, and is uniquely related to activation in somatosensory and motor cortices. First, we established that self-reported patterns of bodily sensation in response to the stimuli were organized by the emotional content of the stimuli: Cluster analysis revealed that self-reports of embodiment, on average, were organized by emotion-related categories such as appetitiveness, disgust, threat, sociality, and sexuality. This was true in a larger online sample of participants, and therefore indicates that there may be some universality to somatosensations related to embodied emotion. Next, we found evidence for our primary hypothesis that primary somatosensory and motor cortex representations are related to self-reported bodily emotion sensation. While the results of the embodied emotion searchlight analysis do not survive multiple comparison correction, we did find evidence for our a priori hypothesis in the uncorrected whole brain data. Furthermore, it is important to recognize that the present study is limited by a small sample size (N = 21) and that the embodiment searchlight results are underpowered because of this. Current work is ongoing to replicate this analysis in a larger sample.

In summary, the relationship between emotion and the body is not purely conceptual: It is supported by physiological responses and perceptual representations. Human emotion is a complex phenomenological experience instantiated by neurobiological and psychological processes that include interactions between the brain and body. This investigation provides evidence that the experience of emotion cannot be isolated from experiences with the world and internal states; instead, emotion emerges from dynamic and distributed representations activated throughout the brain and body. Emotion-related bodily representations may serve to ready an organism for social or survival-related action. Knowledge of these representations may contribute to the biomarker initiative and provide neural targets for emotion regulation in the clinic. The results suggest that therapies targeting bodily sensations could potentially influence emotion by altering somatosensory components of emotion's neural construction.